\documentclass[twocolumn, twocolappendix]{aastex631}
\usepackage{amsmath}
\usepackage{mathrsfs}
\usepackage{graphicx}
\usepackage{xcolor}
\usepackage{dcolumn}
\usepackage{bm}
\usepackage{amssymb}
\usepackage{amsmath}
\usepackage{verbatim}
\usepackage[utf8]{inputenc}
\usepackage{tikz,hyperref}
\usepackage{dcolumn}
\usepackage{multirow}

\shortauthors{Yang, Zhang, Bi, \& Yin}

\begin{document}

\title{Tidal Suppression of Fuzzy Dark Matter Heating in Milky Way Satellite Galaxies}

\author[0009-0005-5375-9437]{Yu-Ming Yang}
\email{yangyuming@ihep.ac.cn}
\affiliation{State Key Laboratory of Particle Astrophysics, Institute of High Energy Physics, Chinese Academy of Sciences, Beijing 100049, China}
\affiliation{School of Physical Sciences, University of Chinese Academy of Sciences, Beijing 100049, China }

\author[0009-0004-1366-1294]{Zhao-Chen Zhang}
\email{zhangzhaochen@ihep.ac.cn}
\affiliation{State Key Laboratory of Particle Astrophysics, Institute of High Energy Physics, Chinese Academy of Sciences, Beijing 100049, China}
\affiliation{School of Physical Sciences, University of Chinese Academy of Sciences, Beijing 100049, China }

\author[0000-0002-5334-9754]{Xiao-Jun Bi}
\email{bixj@ihep.ac.cn}
\affiliation{State Key Laboratory of Particle Astrophysics, Institute of High Energy Physics, Chinese Academy of Sciences, Beijing 100049, China}
\affiliation{School of Physical Sciences, University of Chinese Academy of Sciences, Beijing 100049, China }

\author[0000-0001-6514-5196]{Peng-Fei Yin}
\email{yinpf@ihep.ac.cn}
\affiliation{State Key Laboratory of Particle Astrophysics, Institute of High Energy Physics, Chinese Academy of Sciences, Beijing 100049, China}

\begin{abstract}
Many previous studies have imposed stringent constraints on the particle mass of fuzzy dark matter (FDM) by analyzing observations of Galactic satellite galaxies, which show no significant evidence of the heating effect predicted by FDM. However, these analyses have generally neglected the tidal influence of the Milky Way, which can substantially suppress the FDM-induced heating effect in satellites. This oversight arises from computational challenges of accurately capturing the tidal effects in FDM simulations. In this study, we present a novel simulation framework that, for the first time, enables the simulation of an FDM-stellar system within an observationally motivated gravitational potential of the Milky Way. This framework incorporates the diverse Galactic components, including the gravitational influence of the Large Magellanic Cloud. Using the Fornax dwarf galaxy as a case study, we demonstrate that tidal effects significantly alleviate the tension between observational data and the predicted heating effect for an FDM particle mass of $m_a\sim 10^{-22}$ eV.
\end{abstract}

\keywords{
\href{http://astrothesaurus.org/uat/353}{Dark matter (353)};
\href{http://astrothesaurus.org/uat/416}{Dwarf galaxies(416)};
\href{http://astrothesaurus.org/uat/1880}{Galaxy dark matter halos (1880)}
}

\section{Introduction}
In recent years, many studies have imposed stringent constraints on the mass of fuzzy dark matter (FDM) particles by examining their dynamical heating effect within Milky Way (MW) dwarf galaxies \citep{Marsh:2018zyw, Dalal:2022rmp, Teodori:2025rul}. These studies show that, at very low $m_a$, heating induced by FDM can lead to a substantial expansion of the half-light radius and an increase in stellar velocity dispersion of the dwarf galaxies \citep{Bar_Or_2019, Church_2019, El_Zant_2020, Chiang:2021uvt, Chiang:2022rlx, Dutta_Chowdhury_2021, Dutta_Chowdhury_2023}, which are inconsistent with observational data. However, these analyses neglect the critical influence of Galactic tidal effects. Tidal stripping of the outer Navarro-Frenk-White (NFW)  \citep{Navarro:1995iw}-like region of an FDM halo can weaken interference patterns of the wave function  \citep{Schive:2019rrw}, thereby suppressing gravitational fluctuations arising from soliton oscillation, random walk, and granule evolution. This suppression may significantly suppress the heating effect. 

Accurately capturing tidal effects in FDM simulations is a challenge, primarily due to the large size disparity between the host halo and its subhalo, which necessitates exceptionally high resolution. One direct solution is the zoom-in approach that enhances resolution locally around the subhalo \citep{Chan:2025hhg}. Alternatively, employing a reference frame centered on the subhalo itself can effectively reduce the required simulation volume. While this latter method has been explored, its application has largely been confined to examining tidal effects on isolated solitons \citep{Widmark:2023dec} or FDM subhalos on circular orbits within a simplified point-mass Galactic potential \citep{Schive:2019rrw}. Accurately simulating a realistic system, including both FDM and stars embedded in an observationally motivated Galactic gravitational potential, remains a challenge.

In this study, we present a novel simulation framework formulated within the rest frame of the subhalo, which incorporates a realistic Galactic potential encompassing the DM halo, bulge, stellar discs, and gas discs, as well as the gravitational influence of the Large Magellanic Cloud (LMC). Applying this framework to the Fornax dwarf galaxy with an FDM particle mass of $m_a=10^{-22}$ eV, we find that tidal effects can significantly suppress the heating-induced increases in both the half-light radius and stellar velocity dispersion. Our results greatly alleviate the tension between low values of $m_a$ and observations, thereby relaxing previous stringent constraints of $m_a\gtrsim 10^{-21}$ eV \citep{Teodori:2025rul}. Furthermore, this framework can be readily extended to other Galactic dwarf galaxies, facilitating future studies.

This Letter is organized as follows. Section \ref{Sec2} describes our simulation setup, including the orbit of Fornax, the tidal potential it experiences, and the initialization and evolution of the system. The simulation results are presented in Section \ref{Sec3}, followed by discussions and conclusions in Section \ref{Sec4}. Additional details on the models of the MW and LMC potentials are given in Appendix \ref{App_A}. The treatment of dynamical friction from the Galactic DM halo acting on Fornax is described in Appendix \ref{App_B}, while the derivation of the tidal potential and the details of the system initialization are provided in Appendices \ref{App_C} and \ref{App_D}, respectively.

\section{Simulation Setup\label{Sec2}}
\subsection{Orbits}
We define an inertial reference frame centered at the present-day Galactic center and stationary with respect to its current velocity. Within this frame, the $x$-axis is oriented from the Sun toward the Galactic center, while the $z$-axis points toward the north Galactic pole \citep{vanderMarel:2002kq}. In this coordinate system, the Sun's current position and velocity are $(-8,0,0)$ kpc and $(U_\odot,220\text{ km/s}+V_\odot, W_\odot)$, respectively, where $(U_\odot, V_\odot, W_\odot)=(11.1,12.24,7.25)$ km/s denotes its motion relative to the local standard of rest \citep{Sch_nrich_2010, f2381f32-aadb-3c8a-a473-53334cac15e7}. The current positions and velocities of Fornax and LMC are derived by transforming their observed orbital parameters--heliocentric distance ($D_\odot$), equatorial coordinates ($\alpha$, $\delta$), systemic line-of-sight velocity ($v_\text{sys}$), and Gaia-eDR3 \citep{2021} proper motions ($\mu_{\alpha\star}\equiv\mu_\alpha\cos\delta$, $\mu_\delta$)--into this frame. Detailed values and references are provided in Table \ref{Tab1}.

\begin{table}[htbp]
    \centering
    \caption{Orbital Parameters Provided by Observations}
    \begin{tabular}{cccc}
    \hline
    \hline
    Parameter&Fornax&LMC&Reference\\
    \hline
    $D_\odot$&$139.3\pm2.6$ &$44.59\pm0.09\pm0.54$& 1, 2 \\
    $\alpha$&$39.96667$ &$80.05$& 1, 2 \\
    $\delta$&$-34.51361$ &$-69.30$& 1, 2 \\
    $v_\text{sys}$&$54.1\pm0.5$ &$262.2\pm3.4$& 3, 4 \\
    $\mu_{\alpha\star}$&$0.381\pm0.001$ &$1.910\pm0.020$& 1, 5\\
    $\mu_\delta$&$-0.358\pm0.002$ &$0.229\pm0.047$& 1, 5\\
    \hline
    \end{tabular}
    \label{Tab1}
    \tablecomments{Observed orbital parameters of Fornax and LMC: heliocentric distance $D_\odot$ (kpc), equatorial coordinates $\alpha$ and $\delta$ (deg), systemic line-of-sight velocity $v_\text{sys}$ (km/s), Gaia-eDR3 proper motions $\mu_{\alpha\star}\equiv\mu_\alpha\cos\delta$ and $\mu_\delta$ (mas/yr). References are (1) \cite{Battaglia_2022}; (2) \cite{Pietrzy_ski_2019}; (3) \cite{Battaglia:2006up}; (4) \cite{vanderMarel:2002kq}; (5) \cite{Kallivayalil:2013xb}. }
\end{table}

We evolve the system comprising the MW, the LMC, and Fornax backward in time for $10$ Gyr--approximately the age of Fornax--beginning with the initial conditions above \citep{Pace_2022}. The orbital evolution accounts for the mutual gravitational attraction between the MW and the LMC, as well as their combined influence on Fornax. The gravitational potential of the MW includes contributions from six components \citep{McMillan_2016}: a DM halo, a bulge, two stellar discs, and two gas discs, while the potential of the LMC is modeled as that corresponding to a Hernquist density profile \citep{Erkal_2020}. To account for the effects of dynamical friction exerted by the Galactic halo on both the LMC and Fornax, we employ the Chandrasekhar formula \citep{Chandrasekhar:1943ys}, which is valid given the large orbital scales relative to the FDM de Broglie wavelength \citep{Bar_Or_2019, Lancaster:2019mde}. Further details regarding the potential and friction are provided in Appendices \ref{App_A} and \ref{App_B}, respectively.

\begin{figure}[htbp]
    \centering
    \includegraphics[width=0.9\linewidth]{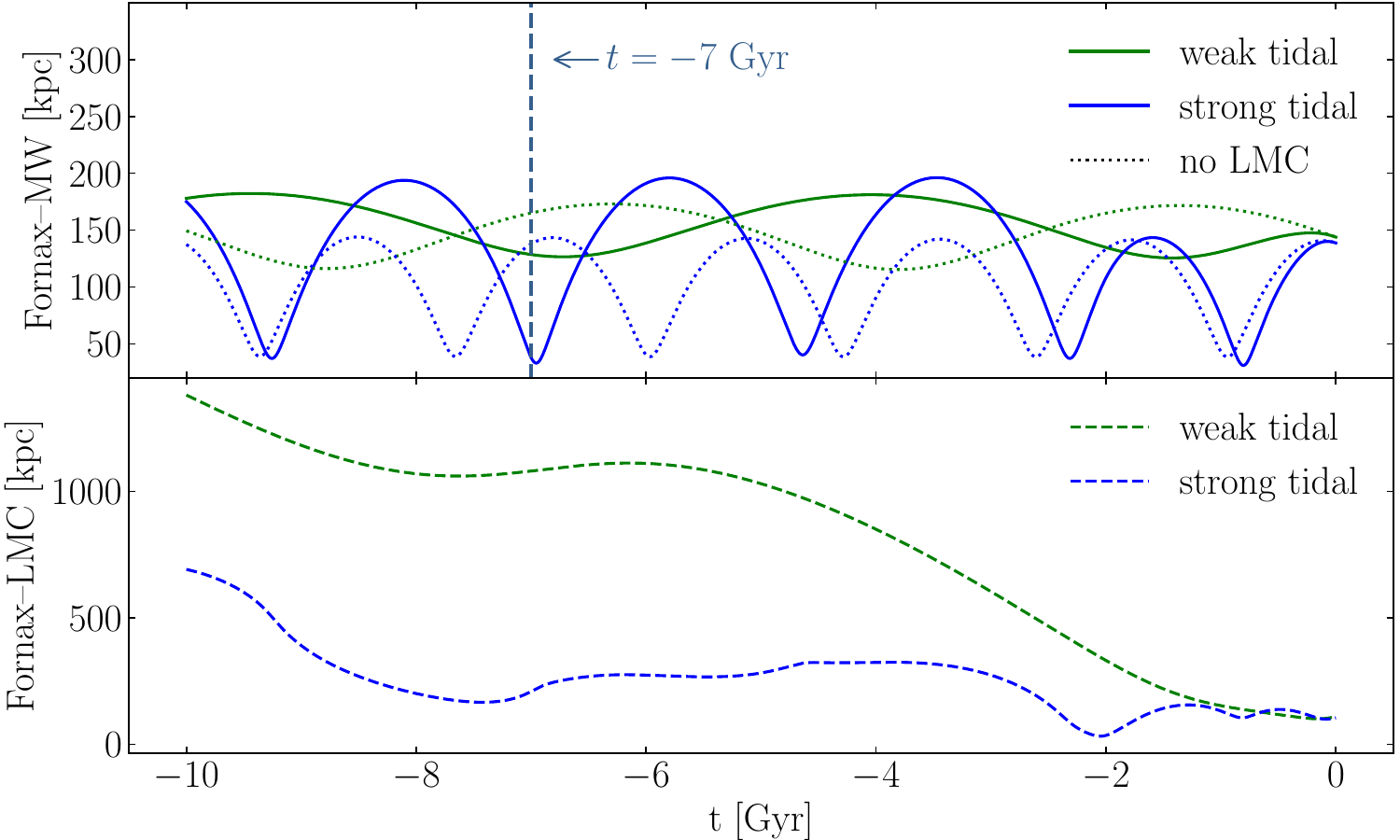}
    \caption{Orbital evolution of Fornax. Top and bottom panels show Fornax’s distance to the centers of the MW and the LMC over time, respectively. Green and blue lines  represent the weak and strong tidal cases, respectively. In the top panel, dotted lines illustrate the scenario excluding the influence of the LMC.}
    \label{orbits}
\end{figure}

We conduct simulations for two orbital scenarios corresponding to weak and strong tidal effects. In each case, the Galactic halo parameter $\rho_{0,h}$ and the observed orbital parameters ($D_{\odot},v_\text{sys},\mu_{\alpha\star}$, and $\mu_\delta$) of Fornax are set to the extreme values within their respective $1\sigma$ range, thereby either minimizing or maximizing tidal effects. The corresponding virial masses of the Galactic halo are $\sim7.34\times10^{11} M_\odot$ and $\sim2.77\times10^{12} M_\odot$, respectively. All other parameters of the MW and the LMC are fixed to their best-fit values. Figure \ref{orbits} shows the evolution of Fornax’s distance to both the MW (upper panel) and the LMC (lower panel). The upper panel also includes a scenario without the influence of the LMC, underscoring its impact on the orbital dynamics of Fornax.

\subsection{Tidal potential}

We define the orbits of Fornax relative to the centers of the MW and the LMC as $\boldsymbol{r}_I(t)=(x_I(t), y_I(t), z_I(t))$, where $I$ represents either the MW or the LMC, and $t\in [-10,0]$ Gyr is the lookback time. Assuming that the soliton center of Fornax follows these orbits, the gravitational influences of the MW and the LMC in the soliton frame can be consistently described by the tidal potential
\begin{equation}
\begin{aligned}
    V^{(I)}_\text{tidal}(\boldsymbol{r}_s,\boldsymbol{r}_I)\simeq &\frac{1}{2}\left[\frac{\partial V_I}{\partial R_I}\frac{(r_{s,x}y_I-r_{s,y}x_I)^2}{(x_I^2+y_I^2)^{3/2}}\right.\\
    &\left.+\frac{\partial^2 V_I}{\partial R_I^2}\frac{(r_{s,x}x_I+r_{s,y}y_I)^2}{x_I^2+y_I^2}+\frac{\partial^2 V_I}{\partial z_I^2}r_{s,z}^2\right.\\
    &\left.+2\frac{\partial^2 V_I}{\partial R_I\partial z_I}\frac{(r_{s,x}x_I+r_{s,y}y_I)r_{s,z}}{(x_I^2+y_I^2)^{1/2}}\right],
\end{aligned}
\label{tidal}
\end{equation}
where $V_I$ denotes the gravitational potential of the MW or LMC, $R_I\equiv (x_I^2+y_I^2)^{1/2}$, and $\boldsymbol{r}_s=(r_{s,x},r_{s,y},r_{s,z})$ represents the coordinate in the soliton frame. A detailed derivation of Equation (\ref{tidal}) is provided in Appendix \ref{App_C}. In our simulations, tidal effects are incorporated by evaluating the tidal potential given in Equation (\ref{tidal}) at Fornax’s instantaneous position $\boldsymbol{r}_I(t)$, and subsequently superimposing the resulting tidal potential to Fornax’s self-gravity.

\begin{figure*}[htbp]
  \centering
  \includegraphics[width=0.364\textwidth]{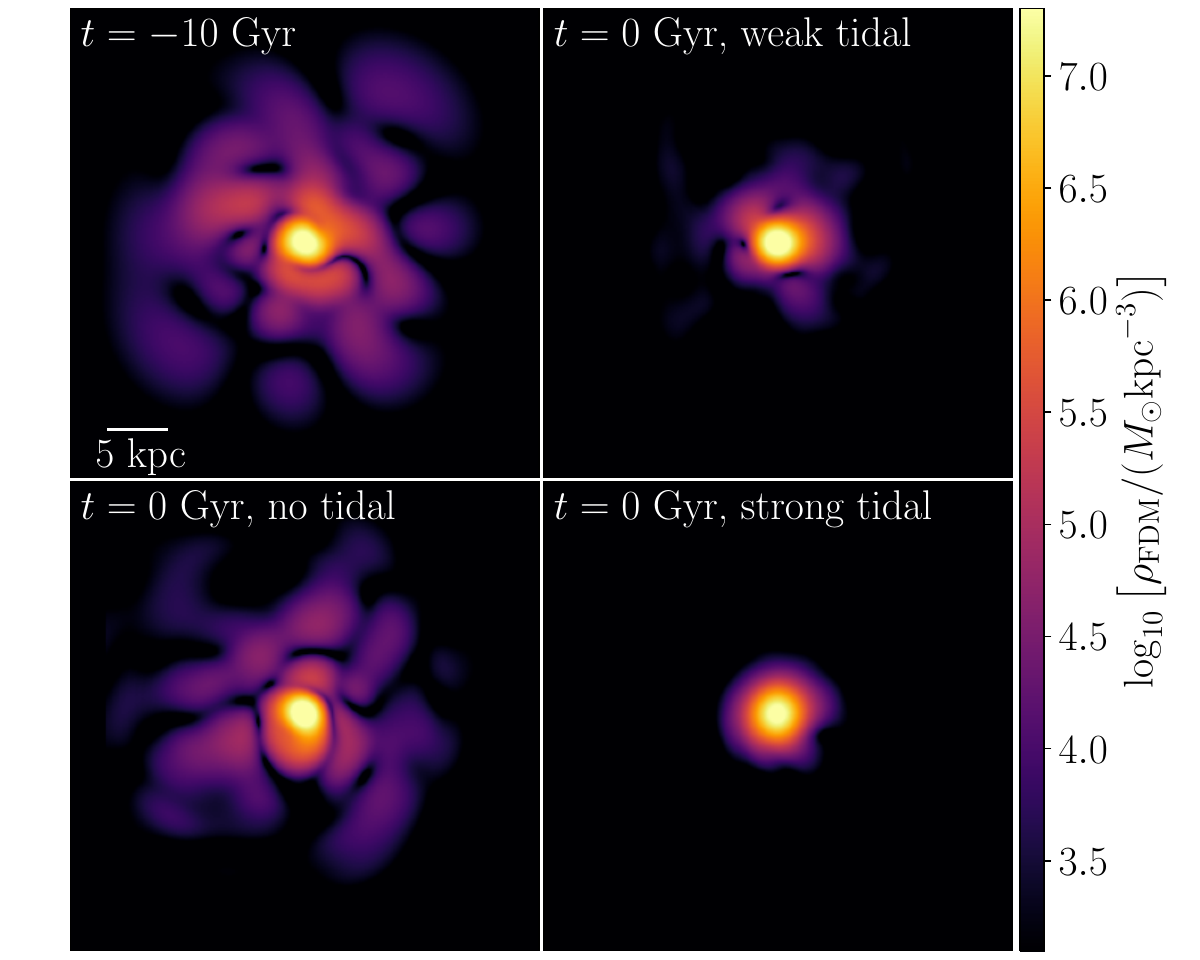}
  \includegraphics[width=0.55\textwidth]{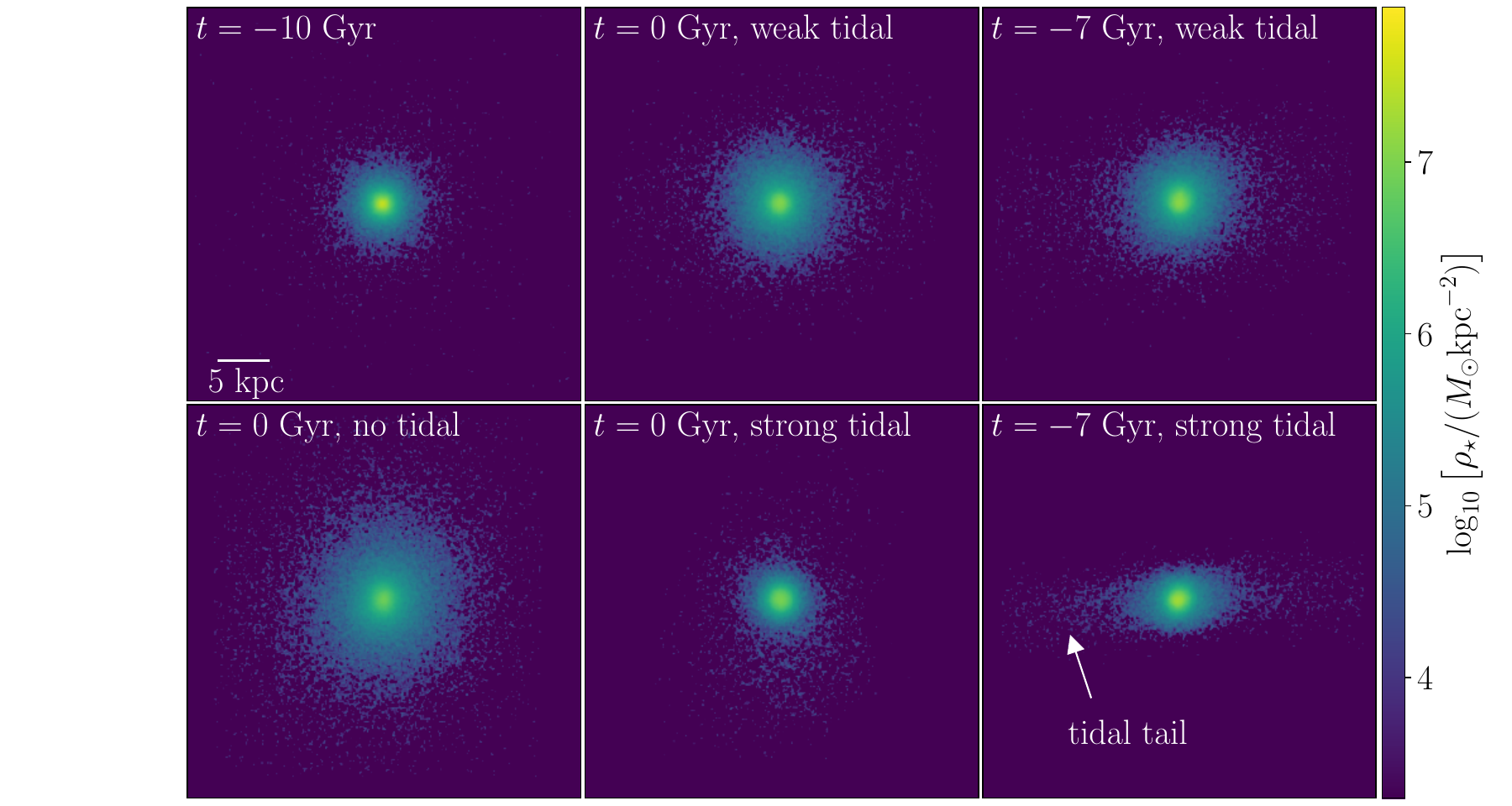}
  \caption{Left: 3D FDM density field. These panels show the 3D FDM density distribution, $\rho_\mathrm{FDM}(t,\boldsymbol{x})=m_a|\psi(t,\boldsymbol{x})|^2$, on a plane centered on the solitonic core. The upper-left panel depicts the initial state at $t = -10$ Gyr, while the remaining panels show the  final states under no, weak, and strong tidal conditions, respectively. Right: 2D stellar density field. The panels display stellar density maps obtained via the particle-mesh method after projecting stellar particle positions onto the $z=0$ plane. The first two columns correspond to the four states shown on the left side, while the last column displays the stellar distributions near pericenter at $t = -7$ Gyr for the weak and strong tidal cases.}
  \label{density_field}
\end{figure*}

\subsection{Initialization and Evolution of the System}

We simulate a two-component system comprising wave-like FDM \citep{hu2000fuzzy, Peebles:2000yy, Marsh:2015xka, Hui:2021tkt, Hui:2016ltb} and particle-like stars. In the non-relativistic limit, FDM is described by an effective wave function $\psi(t,\boldsymbol{x})$ \citep{hu2000fuzzy}, with its corresponding mass density $\rho_\text{FDM}(t,\boldsymbol{x})=m_a|\psi(t,\boldsymbol{x})|^2$ \citep{Hui:2021tkt}, where $\boldsymbol{x}$ denotes coordinates within the simulation box. In contrast, stars are modeled as discrete point masses. The dynamics of these two components are governed by the Schr$\ddot{\text{o}}$dinger equation \citep{Hui:2021tkt, Hui:2016ltb}
\begin{equation}
        i\hbar\frac{\partial\psi}{\partial t}=-\frac{\hbar^2}{2m_a}\nabla^2\psi+m_a V\psi,
\label{Schrodinger}
\end{equation}
and Newton's second law
\begin{equation}
        \frac{d^2\boldsymbol{x}_i}{dt^2}=-\nabla V,\quad i=1,\cdots,N,
\label{Newton}
\end{equation}
respectively. Here, the total gravitational potential $V$ includes both the self-gravity of the FDM and stars ($V_\text{FDM}$ and $V_\text{stars}$), as well as the external tidal potentials arising from the MW and the LMC ($V^{(\text{MW})}_\text{tidal}$ and $V^{(\text{LMC})}_\text{tidal}$). The vector $\boldsymbol{x}_i$ represents the position of the $i$-th star particle within the simulation box.

We initialize the wave function of Fornax's halo at $t=-10$ Gyr using the eigenstate decomposition method \citep{Yavetz:2021pbc, Yang:2024trr, Yang:2024ixt}, which ensures a density profile characterized by a solitonic core surrounded by an NFW-like envelope \citep{Schive:2014dra, Schive:2014hza, Veltmaat:2018dfz, Liao:2024zkj, Mocz:2017wlg, Schwabe:2016rze, Chan:2021bja, Blum:2025aaa}. This choice of initial condition is well motivated, as cosmological simulations indicate that such profiles can form at sufficiently high redshift \citep{Chan:2025hhg, Chiu:2025vng, Schwabe:2021jne}. The stellar component is represented by $N=10^5$ particles, with their initial positions sampled from a Plummer profile \citep{1911MNRAS..71..460P} and their velocities assigned according to the Eddington formula \citep{10.1093/mnras/76.7.572}. We have verified that the stellar distribution remains stable under the combined gravitational potential of the input static FDM profile and the stars. This initialization process involves three free parameters: the FDM halo mass $M_{200}$, the stellar mass $M_\star$, and the initial 2D half-light radius $R_{1/2,\text{initial}}$. Through extensive simulation trials, we adopt $M_{200}=4.0\times 10^8 M_\odot$, $M_\star=5.0\times 10^7 M_\odot$, and $R_{1/2,\text{initial}}=0.7$ kpc, respectively. Further details of the initialization procedure are provided in Appendix \ref{App_D}.

We use \textsc{PyUltraLight} \citep{Edwards:2018ccc}, employing the pseudo-spectral method, to evolve the FDM wave function with a time step of $1$ Myr, following Equation (\ref{Schrodinger}). The simulation box is configured with a resolution of $256^3$ and a half-side length of $L/2=20$ kpc, which is well beyond the halo’s virial radius ($r_{200}\simeq 8.2$ kpc at $t=-10$ Gyr). Periodic boundaries are imposed, and every $5$ Myr the wave function is multiplied by a boundary function $B(r)=\left\{1+\exp\left[-(0.49L-r)/(L/500)\right]\right\}^{-1}$, where $r$ is the distance from the box center, to prevent re-entry of outgoing FDM \citep{Widmark:2023dec}. Stellar particles are evolved according to Equation (\ref{Newton}) using a 4th-order Runge-Kutta integrator with a time step of $0.1$ Myr. Any particles that exit the simulation box are subsequently removed from the system. We have confirmed that the results remain stable under increased resolution and smaller time steps.

At each time step, we efficiently compute the gravitational potential by applying the particle-mesh method--a technique commonly used in N-body simulations \citep{1978AJ.....83..768H, Knebe:2001av, Merz:2004uq, Vogelsberger:2019ynw}--to map stellar particles onto the grid, thereby constructing the stellar density field $\rho_\star(t,\boldsymbol{x})$. This stellar density is then combined with the FDM density $\rho_\text{FDM}(t,\boldsymbol{x})=m_a|\psi(t,\boldsymbol{x})|^2$ to solve the Poisson equation and derive the combined gravitational potential of both FDM and stars. Subsequently, the tidal potentials $V_\text{tidal}^{(\text{MW})}$ and $V_\text{tidal}^{(\text{LMC})}$ are incorporated to yield the total potential. We conduct three sets of simulations corresponding to scenarios with no, weak, and strong tidal effects, based on the previously described orbital configurations of Fornax.

\section{Results\label{Sec3}}
\begin{figure}[htbp]
  \includegraphics[width=\linewidth]{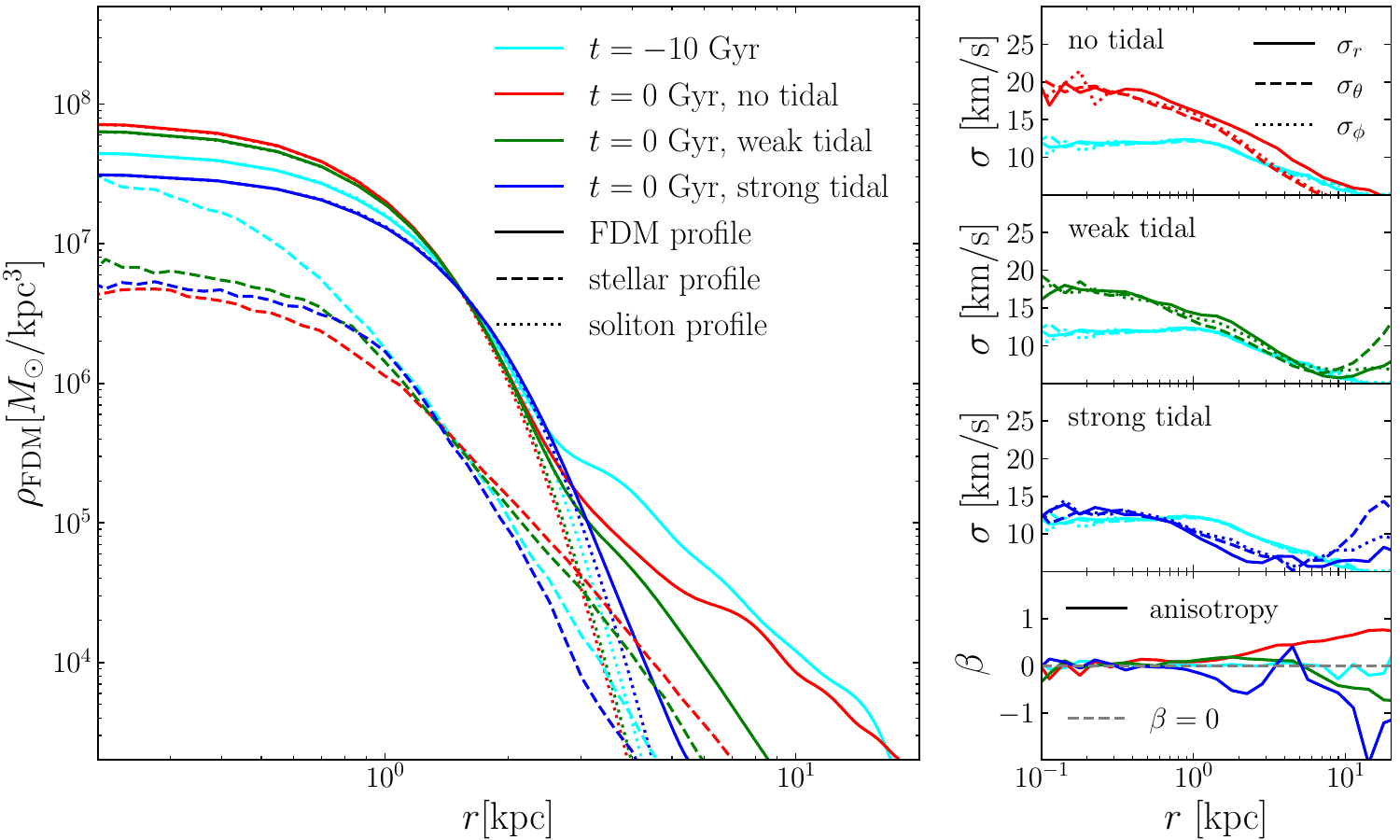}
  \caption{Left: density profiles of FDM and stars. The solid, dotted, and dashed curves represent the spherically averaged FDM density profiles, the corresponding soliton profiles, and the spherically averaged stellar density profiles, respectively.  Right: stellar velocity dispersion and anisotropy profiles. In the top three panels, the solid, dashed, and dotted lines represent the velocity dispersion profiles along the $r$, $\theta$, and $\phi$ directions, respectively. The bottom panel displays the anisotropy parameter as a function of radial distance from the center. The color scheme follows that used in Fig. \ref{result}.}
  \label{profiles}
\end{figure}

The left side of Figure \ref{density_field} displays the 3D FDM density on a plane centered on the solitonic core at the initial time and after $10$ Gyr under no, weak, and strong tidal conditions. The outer NFW-like halo exhibits significant stripping due to tidal forces, with the effect being more pronounced in the stronger tidal scenario. This effect is further illustrated in the left side of Figure \ref{profiles}, where the spherically averaged density profiles (solid lines) demonstrate increased mass loss with stronger tidal strength. In contrast, the central soliton remains intact, consistently following the analytical profile $\rho_\mathrm{soliton}(r)=\rho_c/[1+0.091(r/r_\mathrm{c})^2]^8$ given in the literature \citep{Schive:2014dra, Schive:2014hza} (dotted lines). This consistency aligns well with previous findings \citep{Schive:2019rrw, Du:2018qor}.

The first two columns on the right side of Figure \ref{density_field} show the corresponding 2D stellar densities, obtained by projecting the positions of stellar particles onto the $z = 0$ plane and subsequently mapping them onto the grid using the particle-mesh method. Since Fornax is near apocenter now (Figure \ref{orbits}), tidal tails are not clearly visible. To better illustrate tidal features, the last column presents the stellar densities near pericenter evaluated at $t = -7$ Gyr (see Figure \ref{orbits}) for both the weak and strong tidal cases, with the latter exhibiting significantly more prominent tidal tails.

Comparing the 2D stellar densities at $t = -10$ and $0$ Gyr in the absence of tidal effects (first column panels on the right side of Figure \ref{density_field}), it is evident that the stellar distribution undergoes significant expansion due to dynamical heating. This expansion is notably suppressed in both the weak and strong tidal cases (second column panels), partly because the outer stars are stripped. Moreover, as the outer NFW-like region comprises excited states of the Schr$\ddot{\text{o}}$dinger equation, tidal stripping effectively reduces their number and density, thereby weakening interference patterns with the soliton and among excited states. This reduction diminishes FDM effects, including soliton oscillation, random walk, and granule fluctuations, thereby reducing the heating of both inner and outer stars. 

The left panel of Figure \ref{profiles} shows the spherically averaged 3D stellar density profiles (dashed lines), computed relative to the stellar density peak. As expected, stronger tidal strength leads to more pronounced stripping in the outer region. At $t=0$ Gyr, the higher inner stellar density in the weak tidal case, compared to the no tidal case, reflects the suppression of dynamical heating. Conversely, in the strong tidal case, the inner stellar density is lower due to tidal forces sufficiently intense to pull even the inner stars outward. In the weak and strong tidal cases, the remaining stellar mass within $2$ kpc at $t=0$ Gyr is $3.1\times 10^7M_\odot$ and $2.8\times 10^7M_\odot$, respectively, which are consistent with the values reported in the literature \citep{McConnachie_2012, de_Boer_2012}.

\begin{figure}[htbp]
    \centering
    \includegraphics[width=\linewidth]{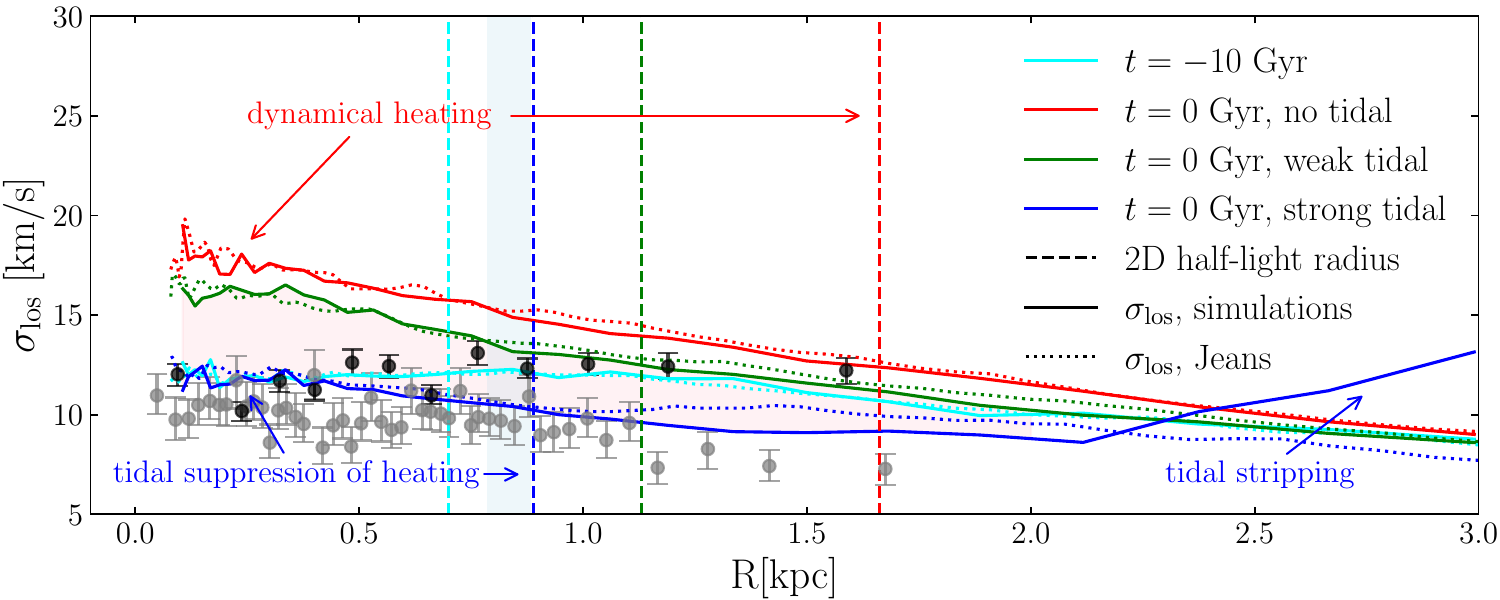}
    \caption{Tidal suppression of FDM-induced dynamical heating in Fornax. Vertical dashed lines indicate the 2D half-light radius. Solid and dotted lines represent the line-of-sight velocity dispersion profiles obtained from simulations and by solving the Jeans equation, respectively. Cyan lines denote the initial state ($t = -10$ Gyr), while red, green, and blue lines represent the final states ($t = 0$ Gyr) under conditions of no, weak, and strong tidal effects, respectively. The pink shaded area between the green and blue solid lines reflects the uncertainties arising from the Galactic halo density parameter $\rho_{0,h}$ and the Fornax's orbital parameters. The vertical light blue band indicates the $1\sigma$ uncertainty range of the 2D half-light radius derived in \cite{Wang_2019}, due to the uncertainty in Fornax’s distance. Gray and black points with error bars represent velocity dispersion data from \cite{Walker:2009zp} and \cite{Breddels:2013dga}, respectively.}
    \label{result}
\end{figure}

The tidal suppression of dynamical heating directly reduces the growth of the half-light radius. In Figure \ref{result}, vertical dashed lines show the 2D half-light radii under different scenarios. Comparing the vertical cyan and red dashed lines, the no tidal case shows substantial radius growth, well beyond the observed value (vertical light blue band) \citep{Wang_2019}. This discrepancy has been a key factor in previous studies that have ruled out $m_a\lesssim 10^{-21}$ eV \citep{Teodori:2025rul}. However, the inclusion of tidal effects substantially suppresses this growth. Notably, in the strong tidal case, the half-light radius can fall within the range of observational uncertainties even for $m_a = 10^{-22}$ eV.

The reduced growth of the stellar velocity dispersion further supports the suppression of dynamical heating by tidal effects. The right side of Figure \ref{profiles} shows velocity dispersion profiles along three directions and the radial variation of the anisotropy parameter in different scenarios. In the absence of tidal effects (top panel), dynamical heating significantly enhances velocity dispersion in all directions, and produces positive anisotropy at large radii, consistent with previous studies \citep{Dutta_Chowdhury_2023}. In contrast, when tidal effects are included (second and third panels from the top), the increase in dispersion is notably suppressed within a radius of $10$ kpc. Beyond this radius, in the $10-20$ kpc region, the rising radial dispersion $\sigma_r$ reflects stars that have been tidally stripped and are escaping Fornax's gravitational potential. The even larger increase in the tangential components $\sigma_\theta$ and $\sigma_\phi$ indicates the tidal-induced rotation, resulting in a strongly negative anisotropy at large radii (bottom panel).

Since only the line-of-sight velocity dispersion is observable, Figure \ref{result} compares simulated values, as observed from the Earth (solid lines) under different scenarios, with observational data \citep{Walker:2009zp, Breddels:2013dga}. The initial velocity dispersion aligns with the observational data from \citep{Breddels:2013dga}. However, in the absence of tidal effects, this dispersion increases significantly, especially in the inner region, surpassing observed values and prompting previous studies to impose stringent constraints on $m_a$. Including tidal effects significantly reduces the velocity dispersion, with the strong tidal case even resulting in values that drop below the initial values due to a decrease in the dynamical mass. The pink shaded region between the green and blue solid lines in this figure represents the uncertainties arising from the Galactic halo density parameter $\rho_{0,h}$ and Fornax's orbital parameters. When tidal effects are included, the simulated results show no significant deviation from the observations within these theoretical uncertainties. In particular, under strong tidal conditions, the simulation closely matches the observational data presented in \cite{Breddels:2013dga}, as shown by the gray data points.

Many previous studies have constrained $m_a$ through Jeans analysis \citep{Chen:2016unw, Luu:2018afg, Hayashi:2021xxu, DeMartino:2023cgg, Zimmermann:2024xvd}, but its validity in the presence of both dynamical heating and tidal effects has not been widely tested. To address this, we solve the Jeans equation using the FDM and stellar density profiles, along with the anisotropy parameter profiles obtained from our simulations. The resulting line-of-sight velocity dispersions (dotted lines in Figure \ref{result}) agree well with the simulation results. This consistency holds except beyond a radius of 2 kpc in the strong tidal case, where discrepancies emerge due to the removal of tidally stripped stars from equilibrium.
\section{Discussions and conclusions\label{Sec4}}

We have presented the first simulation of a dwarf galaxy composed of FDM and stars evolving under the combined tidal influence of the MW and the LMC. A limitation of our simulation framework is the omission of the influence of the Galactic FDM wave function, which would necessitate zoom-in simulations. However, such simulations remain technically challenging at $m_a\gtrsim 10^{-22}$ eV \citep{Chan:2025hhg}. As noted in \citep{Chan:2025hhg}, tidal stripping of FDM subhalos leads to smaller granules from the host halo dominating their outskirts. Nonetheless, since heating efficiency scales with granule volume \citep{Bar_Or_2019}, the impact of these smaller structures on our analysis is expected to be minimal.

One of the key conclusions from our simulations is that incorporating tidal effects can effectively alleviate the tension between an FDM particle mass of $10^{-22}$ eV and the observational data. This highlights the crucial role of tidal interactions in constraining FDM using Galactic satellites, as neglecting them can lead to order-of-magnitude deviations. Moreover, our simulation framework is readily extendable to other dwarf galaxies. For instance, some dwarf galaxies \citep{Pace_2022}, such as Segue 1 and Segue 2, reside much closer to the Galactic center than Fornax, where tidal effects are even more pronounced, suggesting that previous constraints based on these galaxies (such as in \cite{Dalal:2022rmp}) may be overly stringent. 

It is worth noting that an FDM particle mass of $10^{-22}$ eV may still be in tension with other observational constraints \citep{Eberhardt:2025caq}, such as those derived from the subhalo mass function \citep{DES:2020fxi}, the Ly$\alpha$ forest \citep{Rogers:2020ltq}, and strong lensing observations \citep{Laroche:2022pjm}. Further investigation into these constraints is needed.
\section*{acknowledgments}
This work is supported by the National Natural Science Foundation of China under grant No. 12447105.

\appendix

\section{Density profiles of the MW and the LMC \label{App_A}}
The density structure of the MW is modeled following the prescription of \cite{McMillan_2016}, which comprises six distinct components. These include a Navarro-Frenk-White (NFW) dark matter (DM) halo,
\begin{equation}
    \rho_{h}(r)=\frac{\rho_{0,h}}{\frac{r}{r_h}\left(1+\frac{r}{r_h}\right)^2},
    \label{halo}
\end{equation}
a central bulge,
\begin{equation}
    \rho_b(R,z)=\frac{\rho_{0,b}}{\left[1+\frac{\sqrt{R^2+(z/q)^2}}{r_0}\right]^\alpha}\exp\left[-\frac{R^2+(z/q)^2}{r^2_\mathrm{cut}}\right],
\end{equation}
a thin and a thick stellar disc,
\begin{equation}
    \rho^s_{d,I}(R,z)=\frac{\Sigma_{0,I}}{2z_{d,I}}\exp\left(-\frac{|z|}{z_{d,I}}-\frac{R}{R_{d,I}}\right),
\end{equation}
where $I=\mathrm{thin, thick}$, and HI and molecular (H$_2$) gas discs,
\begin{equation}
    \rho^g_{d,J}(R,z)=\frac{\Sigma_{0,J}}{4z_{d,J}}\exp\left(-\frac{R_{m,J}}{R}-\frac{R}{R_{d,J}}\right)\mathrm{sech}^2\left(\frac{z}{2z_{d,J}}\right),
\end{equation}
where $J=\mathrm{HI, H}_2$. In the above expressions, $x$, $y$, and $z$ denote Cartesian coordinates in the Galactic reference frame, with $r = \sqrt{x^2 + y^2 + z^2}$ and $R = \sqrt{x^2 + y^2}$ representing the three-dimensional and projected radial distances from the Galactic center, respectively. All parameters are fixed to the global best-fit values reported in \cite{McMillan_2016}, except for the DM halo density parameter $\rho_{0,h}$, which is adjusted to account for weak or strong tidal scenarios.

For the LMC, following the approach of \cite{Erkal_2020} and \cite{Pace_2022}, we model its density distribution using a Hernquist profile
\begin{equation}
    \rho_\mathrm{LMC}(r)=\frac{M_\mathrm{LMC}}{2\pi a^3}\frac{1}{\frac{r}{a}\left(1+\frac{r}{a}\right)^3},
\end{equation}
where $r$ is the three-dimensional distance from the LMC center. The total mass of the LMC is set to the best-fit value from \cite{Erkal_2019}, $M_\mathrm{LMC}=1.38\times 10^{11} M_\odot$. The scale radius $a$ is chosen to ensure that the LMC has a circular velocity of $V_\mathrm{circ}=91.7$ km/s at a galactocentric distance of $8.7$ kpc \citep{vanderMarel:2013jza}. All density parameters adopted for the MW and the LMC in our simulations are summarized in Table \ref{Tab2}.

\begin{table}[htbp]
    \centering
    \caption{Density parameters of the MW and the LMC}
    \begin{tabular}{ccc}
    \hline
    \hline
    \multirow{3}{*}{DM halo}&\multirow{2}{*}{$\rho_{0,h}$}&$5.3\times 10^6$ (weak tidal)\\
                & &$1.59\times 10^7$ (strong tidal)\\
                &$r_h$& $19.6$\\
    \hline
    \multirow{5}{*}{Bulge}&$\rho_{0,b}$&$9.84\times 10^{10}$\\
                &$\alpha$&$1.8$\\
                &$q$&$0.5$\\
                &$r_0$&$0.075$\\
                &$r_\mathrm{cut}$&$2.1$\\
    \hline
    \multirow{3}{*}{Thin stellar disc}&$\Sigma_{0,\mathrm{thin}}$&$8.96\times 10^8$\\
                &$z_{d,\mathrm{thin}}$&$0.3$\\
                &$R_{d,\mathrm{thin}}$&$2.5$\\
    \hline
    \multirow{3}{*}{Thick stellar disc}&$\Sigma_{0,\mathrm{thick}}$&$1.83\times 10^8$\\
                &$z_{d,\mathrm{thick}}$&$0.9$\\
                &$R_{d,\mathrm{thick}}$&$3.02$\\
    \hline
    \multirow{4}{*}{HI disc}&$\Sigma_{0,\mathrm{HI}}$&$5.31\times 10^7$\\
                &$z_{d,\mathrm{HI}}$&$0.085$\\
                &$R_{m,\mathrm{HI}}$&$4$\\
                &$R_{d,\mathrm{HI}}$&$7$\\
    \hline
    \multirow{4}{*}{H$_2$ disc}&$\Sigma_{0,\mathrm{H_2}}$&$2.18\times 10^9$\\
                &$z_{d,\mathrm{H_2}}$&$0.045$\\
                &$R_{m,\mathrm{H_2}}$&$12$\\
                &$R_{d,\mathrm{H_2}}$&$1.5$\\
    \hline
    \multirow{2}{*}{LMC}&$M_\mathrm{LMC}$&$1.38\times 10^{11}$\\
                &$a$&$16.1$\\
    \hline
    \end{tabular}
    \label{Tab2}
    \tablecomments{Density parameters of the Galactic DM halo $\rho_{0,h}$ ($M_\odot/\mathrm{kpc}^{-3}$) and $r_h$ (kpc); Galactic bulge $\rho_{0,b}$ ($M_\odot/\mathrm{kpc}^{-3}$), $\alpha$, $q$, $r_0$ (kpc), and $r_\mathrm{cut}$ (kpc); thin and thick stellar discs $\Sigma_{0,\mathrm{thin}/\mathrm{thick}}$ ($M_\odot/\mathrm{kpc}^{-2}$), $z_{d,\mathrm{thin}/\mathrm{thick}}$ (kpc), and $R_{d,\mathrm{thin}/\mathrm{thick}}$ (kpc); HI and H$_2$ discs $\Sigma_{0,\mathrm{HI}/\mathrm{H_2}}$ ($M_\odot/\mathrm{kpc}^{-2}$), $z_{d,\mathrm{HI}/\mathrm{H_2}}$ (kpc), $R_{m,\mathrm{HI}/\mathrm{H_2}}$ (kpc), and $R_{d,\mathrm{HI}/\mathrm{H_2}}$ (kpc); the LMC $M_\mathrm{LMC}$ ($M_\odot$) and $a$ (kpc).}
\end{table}
\section{Dynamical friction \label{App_B}}
To accurately determine the orbits of the LMC and Fornax within the MW, it is essential to account for the dynamical friction exerted by the Galactic halo on both systems. In this study, we consider FDM with a particle mass of $m_a=10^{-22}$ eV. Given the characteristic velocity of FDM particles in the MW, $v_\mathrm{typ} \simeq 220$ km/s, their corresponding de Broglie wavelength is approximately $0.5$ kpc, significantly smaller than the orbital scales of the LMC and Fornax. Under these conditions, the dynamical friction force in a FDM halo can be well approximated using the Chandrasekhar formula \citep{Chandrasekhar:1943ys, Bar_Or_2019, Lancaster:2019mde}, which is commonly applied in the cold dark matter (CDM) scenario 
\begin{equation}
    \boldsymbol{F}_\mathrm{DF}=-\frac{4\pi G^2 M_\mathrm{sat} \rho_h(r)\ln\Lambda}{v^2}\left[\mathrm{erf}(X)-\frac{2X}{\sqrt{\pi}}\mathrm{e}^{-X^2}\right]\frac{\boldsymbol{v}}{v},
    \label{friction}
\end{equation}
where $G$ is the gravitational constant, and $M_\mathrm{sat}$, $r$, and $\boldsymbol{v}$ represent the total mass, distance to the Galactic center, and the instantaneous velocity of the LMC or Fornax along their orbits, respectively. The dimensionless parameter $X\equiv v/[\sqrt{2}\sigma(r)]$ depends on the local velocity dispersion $\sigma(r)$ of the Galactic halo. We adopt the analytic approximation which is accurate to within $1\%$ over the range $r\in [0.01r_h,100r_h]$ \citep{Zentner:2003yd} 
\begin{equation}
    \sigma(r)\simeq V_\mathrm{max}\frac{1.4393 (r/r_h)^{0.354}}{1+1.1756(r/r_h)^{0.725}},
\end{equation}
where $V_\mathrm{max}$ is the maximum circular velocity under the Galactic halo potential. The term $\Lambda$ in Eq. \ref{friction} is defined as $\Lambda\equiv b_\mathrm{max}/b_\mathrm{min}$, with $b_\mathrm{max}=r$. Under the point-mass satellite approximation in the FDM framework, the minimum impact parameter is commonly set to $b_\mathrm{min}=\lambda/(4\pi)$, where $\lambda$ is the FDM particle’s de Broglie wavelength \citep{Bar_Or_2019}. However, both the LMC and Fornax possess finite physical sizes much larger than $\lambda/(4\pi)$. Therefore, we adopt $b_\mathrm{min}=4.8$ kpc \citep{Besla:2007kf} for the LMC and $b_\mathrm{min}=1$ kpc for Fornax, respectively. This choice is well justified, as the logarithmic dependence of $\Lambda$, which is typically much greater than unity, ensures that moderate variations in $b_\mathrm{min}$ have a negligible impact on the calculated dynamical friction force.
\section{Derivation of the tidal potential \label{App_C}}
As a good approximation, we assume that the soliton center moves along the orbit determined in the main text, governed by the combined gravitational potential of the MW and the LMC. That is, the position of the soliton center, denoted by $\boldsymbol{r}^{(s)}$, satisfies 
\begin{equation}
    \frac{d^2\boldsymbol{r}^{(s)}}{dt^2}=-\nabla_{\boldsymbol{r}^{(s)}}\left[V^{(\mathrm{MW})}+V^{(\mathrm{LMC})}\right],
\end{equation}
where $\nabla_{\boldsymbol{r}^{(s)}}$ denotes the gradient evaluated at the position $\boldsymbol{r}^{(s)}$. We now consider a test particle in Fornax subjected to the gravitational forces from the MW, the LMC, and Fornax itself. Its motion obeys
\begin{equation}
    \frac{d^2\boldsymbol{r}_t}{dt^2}=-\nabla_{\boldsymbol{r}_t}\left[V^{(\mathrm{MW})}+V^{(\mathrm{LMC})}\right]+\boldsymbol{F}_\mathrm{Fornax}.
\end{equation}
where $\boldsymbol{r}_t$ is the position of the test particle, and $\boldsymbol{F}_\mathrm{Fornax}$ represents the gravitational force from Fornax. The relative motion of the test particle with respect to the soliton center of Fornax is then given by
\begin{equation}
\begin{aligned}
    \frac{d^2\boldsymbol{r}_s}{dt^2}&\equiv\frac{d^2}{dt^2}(\boldsymbol{r}_t-\boldsymbol{r}^{(s)})\\
    &=-\left[\nabla_{\boldsymbol{r}_t}V^{(\mathrm{MW})}-\nabla_{\boldsymbol{r}^{(s)}}V^{(\mathrm{MW})}\right]\\
    &\quad-\left[\nabla_{\boldsymbol{r}_t}V^{(\mathrm{LMC})}-\nabla_{\boldsymbol{r}^{(s)}}V^{(\mathrm{LMC})}\right]+\boldsymbol{F}_\mathrm{Fornax}\\
    &\equiv \boldsymbol{F}^{(\mathrm{MW})}_\mathrm{tidal}+\boldsymbol{F}^{(\mathrm{LMC})}_\mathrm{tidal}+\boldsymbol{F}_\mathrm{Fornax}.
\end{aligned}
\end{equation}
Taking $\boldsymbol{F}^{(\mathrm{MW})}_\mathrm{tidal}$ as an example, in the limit where $r_s \ll r^{(s)}$, it can be approximated as
\begin{equation}
\begin{aligned}
    \boldsymbol{F}^{(\mathrm{MW})}_\mathrm{tidal}(\boldsymbol{r}_s,\boldsymbol{r}^{(s)})&\simeq -\left[\boldsymbol{r}_t-\boldsymbol{r}^{(s)}\right]\cdot\nabla_{\boldsymbol{r}^{(s)}}\left[\nabla_{\boldsymbol{r}^{(s)}}V^{(\mathrm{MW})}\right]\\
    &=-\boldsymbol{r}_s\cdot\nabla_{\boldsymbol{r}^{(s)}}\left[\nabla_{\boldsymbol{r}^{(s)}}V^{(\mathrm{MW})}\right].
\end{aligned}
\end{equation}
Consequently, the corresponding tidal potential at the position of the test particle is given by
\begin{equation}
\begin{aligned}
    V^{(\mathrm{MW})}_\mathrm{tidal}(\boldsymbol{r}_s,\boldsymbol{r}^{(s)})&=-\int_0^{\boldsymbol{r}_s}\boldsymbol{F}^{(\mathrm{MW})}_\mathrm{tidal}(\boldsymbol{r}^\prime_s,\boldsymbol{r}^{(s)})\cdot d\boldsymbol{r}^\prime_s\\
    &=\frac{1}{2}\left[\boldsymbol{r}_s\cdot\nabla_{\boldsymbol{r}^{(s)}}\right]^2V^{(\mathrm{MW})}.
\end{aligned}
\end{equation}
Since all the MW density components adopted in our model are axisymmetric, it is convenient to evaluate the above expression in cylindrical coordinates. By relabeling $\boldsymbol{r}^{(s)}$ as $\boldsymbol{r}_{(\mathrm{MW})}$, one can straightforwardly derive the expression presented in Equation (\ref{tidal}) of the main text for the case $I=\mathrm{MW}$. Applying the same procedure to $\boldsymbol{F}^{(\mathrm{LMC})}_\mathrm{tidal}$ yields an analogous result, confirming that the unified form of the tidal potential given in Equation (\ref{tidal}) can be applied to both the MW and the LMC.
\section{Initialization of the system \label{App_D}}
We initialize the wave function based on an input density profile composed of a solitonic core connected to an outer NFW envelope, consistent with results from cosmological simulations \citep{Schive:2014dra, Schive:2014hza, Chan:2025hhg, Chiu:2025vng, Schwabe:2021jne}
\begin{equation}
    \rho_\mathrm{in}(r)=\left\{\begin{aligned}
        &\frac{\rho_c}{\left[1+0.091(r/r_c)^2\right]^8},\quad r<kr_c\\
        &\frac{\rho_s}{(r/r_s)\left(1+r/r_s\right)^2},\quad r\geq kr_c,
        \end{aligned}\right.
        \label{input_profile}
\end{equation}
where the four parameters $\rho_c$, $r_c$, $\rho_s$, and $r_s$ denote the density and radius of the solitonic core, and the characteristic density and scale radius of the NFW envelope, respectively. The transition between the soliton and NFW regions occurs at $r = k r_c$, with $k$ being a parameter that characterizes the matching point.

Given a virial mass $M_{200}$ (or, equivalently, a virial radius $r_{200}$), we determine the five parameters in Equation (\ref{input_profile}) as follows. First, we compute the concentration parameter $c$ of the NFW envelope using the mass–concentration–redshift relation from \cite{Ludlow:2016ifl}, which allows us to fix $r_s = r_{200}/c$. In this process, the cosmological parameters are taken as $\Omega_m=0.3,\Omega_\Lambda=0.7$, and $H_0=67.36$ km/s/Mpc. Next, we incorporate constraints from cosmological simulations, which suggest a statistical correlation between the soliton mass and the virial mass \citep{Schive:2014hza, Veltmaat:2018dfz, Liao:2024zkj}. To account for the scatter observed in this relation \citep{Mocz:2017wlg, Schwabe:2016rze, Chan:2021bja, Blum:2025aaa}, we denote the best-estimate soliton mass corresponding to a given $M_{200}$ as $M_{c,\mathrm{best}}$, and assume that $\ln(M_c/M_\odot)$ follows a Gaussian distribution $p_1(M_c)$ centered on $\ln(M_{c,\mathrm{best}}/M_\odot)$ with a variance of $\ln 2$. Similarly, since the transition parameter $k$ varies across different studies  \citep{Dutta_Chowdhury_2021, Mocz:2017wlg, Chiang:2021uvt, Blum:2025aaa}, we model $k$ as a Gaussian-distributed parameter $p_2(k)$ with a mean of 3 and variance 0.5. We then scan over a range of possible soliton masses $M_c$. For a given value of  $M_c=2.78\rho_cr_c^3$, combined with the relation $\rho_c=1.95\times 10^7 M_\odot \text{kpc}^{-3}\left(m_a/10^{-22}\text{eV}\right)^{-2}\left(r_c/\text{kpc}\right)^{-4}$, we can uniquely determine $\rho_c$ and $r_c$. The parameters $k$ and $r_s$ are then obtained by enforcing continuity of the density profile at $r = k r_c$ and requiring that the total mass integrates to the given virial mass, $4\pi\int_0^{r_{200}}\rho_\mathrm{in}(r)r^2dr=M_{200}$. Finally, among all combinations of parameters explored, we adopt the set that maximizes the product of the probabilities $p_1(M_c)$ and $p_2(k)$.

The initial stellar distribution is modeled with a Plummer profile \citep{1911MNRAS..71..460P}
\begin{equation}
    \rho_\star(r)=\frac{3M_\star}{4\pi R_{1/2,\mathrm{initial}}^3}\left(1+\frac{r^2}{R_{1/2,\mathrm{initial}}^2}\right)^{-5/2},
\end{equation}
where $M_\star$ is the total stellar mass and $R_{1/2,\mathrm{initial}}$ is the initial 2D half-light radius. The initial FDM halo wave function is constructed as a linear combination of eigenstates, $\psi(0,\mathbf{x})=\sum_{nlm}|a_{nl}|e^{i\phi_{nlm}}\Psi_{nlm}(\mathbf{x})$ \citep{Yavetz:2021pbc}, where $\Psi_{nlm}(\mathbf{x})$ are the eigenfunctions of the time-independent Schr$\ddot{\text{o}}$dinger equation under the combined gravitational potential generated by both the input FDM and stellar density profiles. 
We retain only bound states with eigenenergies below that of a circular orbit at the virial radius, and solve them using the shooting method \citep{Yang:2024ixt}. The phases $\phi_{nlm}$ are randomly assigned, while the amplitudes $|a_{nl}|$ are determined by fitting the output density profile, $\rho_\mathrm{out}(r)=\frac{m_a}{4\pi}\sum_{nl}(2l+1)|a_{nl}|^2R^2_{nl}(r)$, to the input profile $\rho_\mathrm{in}$ using non-negative least squares method. This approach significantly accelerates the fitting process due to the linear dependence of $\rho_\mathrm{out}$ on $|a_{nl}|^2$. The initial bulk velocity of the halo is computed following \cite{Yang:2024trr} and removed via a Galilean boost. For more details on the eigenstate solutions and related methodology, see \citep{Yavetz:2021pbc, Yang:2024trr, Yang:2024ixt}. For the stellar particles, as described in the main text, their positions are sampled according to the Plummer profile \citep{1911MNRAS..71..460P}, while their velocities are generated using the Eddington formula \citep{10.1093/mnras/76.7.572}, which takes into account the combined gravitational potential from both the FDM and stellar components.
\bibliography{Refs}
\bibliographystyle{aasjournal}

\end{document}